\documentclass[useAMS,usenatbib]{mn2e}
\usepackage{epsfig,graphics}

\def\Re{\mbox{$R_{\rm eff}$}}

\def\Msun{\mbox{$M_\odot$}}

\def\ML{\mbox{$M/L$}}
\def\Yst{\mbox{$\Upsilon_{*}$}}

\def\mst{\mbox{$M_{\star}$}}

\def\lsim{\mathrel{\rlap{\lower3.5pt\hbox{\hskip0.5pt$\sim$}}
    \raise0.5pt\hbox{$<$}}}                
\def\gsim{~\rlap{$>$}{\lower 1.0ex\hbox{$\sim$}}}

\def\Msun{\mbox{$M_\odot$}}

\def\lsim{\mathrel{\rlap{\lower3.5pt\hbox{\hskip0.5pt$\sim$}}
    \raise0.5pt\hbox{$<$}}}
\def\gsim{~\rlap{$>$}{\lower 1.0ex\hbox{$\sim$}}}

\def\sig{\mbox{$\sigma$}}
\def\sigc{\mbox{$\sigma_{0}$}}
\def\Re{\mbox{$R_{\rm eff}$}}

\def\mst{\mbox{$M_{*}$}}

\def\gZ{\mbox{$\nabla_{\rm Z}$}}
\def\gage{\mbox{$\nabla_{\rm age}$}}
\def\gML{\mbox{$\nabla_{\rm \tiny \Yst}$}}

\def\ggr{\mbox{$\nabla_{\rm g-r}$}}

\title[M/L gradients]{Stellar mass-to-light ratio gradients in galaxies: correlations with mass.}

\author[Tortora et al.]{\noindent
C.~Tortora$^{1}$\thanks{E-mail: ctortora@physik.uzh.ch},
N.R.~Napolitano$^{2}$, A.J.~Romanowsky$^{3}$, Ph. Jetzer$^{1}$,
V.F. Cardone$^{4}$ \and M.~Capaccioli$^{5}$
\\~\\
$^1$ Universit$\ddot{a}$t Z$\ddot{u}$rich, Institut f$\ddot{u}$r
Theoretische Physik, Winterthurerstrasse 190,
CH-8057, Z$\ddot{u}$rich, Switzerland\\
$^2$ INAF -- Osservatorio Astronomico di Capodimonte, Salita
Moiariello, 16, 80131 - Napoli, Italy\\
$^3$ UCO/Lick Observatory, University of California, Santa Cruz,
CA 95064, USA \\
$^4$ INAF - Osservatorio Astronomico di Roma -
via di Frascati, 33, 00040 - Monte Porzio Catone (Roma), Italy \\
$^5$ Dipartimento di Scienze Fisiche, Universit\`{a} di Napoli
Federico II, Compl. Univ. Monte S. Angelo, 80126 - Napoli, Italy}

\begin{document}
\date{Accepted  Received }
\pagerange{\pageref{firstpage}--\pageref{lastpage}} \pubyear{xxxx}
\maketitle

\label{firstpage}

\begin{abstract}
We analyze the stellar mass-to-light ratio (\ML) gradients in a
large sample of local galaxies taken from the Sloan Digital Sky
Survey, spanning a wide range of stellar masses and morphological
types. As suggested by the well known relationship between
mass-to-light (\ML) ratios and colors, we show that \ML\ gradients
are strongly correlated with colour gradients, which we trace to
the effects of age variations. Stellar \ML\ gradients generally
follow patterns of variation with stellar mass and galaxy type
that were previous found for colour and metallicty gradients. In
late-type galaxies \ML\ gradients are negative, steepening with
increasing mass. In early-type galaxies \ML\ gradients are
shallower while presenting a two-fold trend: they decrease with
mass up to a characteristic mass of $\mst \sim 10^{10.3} \Msun$
and increase at larger masses. We compare our findings with other
analyses and discuss some implications for galaxy formation and
for dark matter estimates.
\end{abstract}


\section{Introduction}\label{sec:intro}

Colour and absorption line gradients are markers of stellar
population variations within galaxies, providing important clues
to galaxy evolution (e.g. \citealt{Spolaor+09}; \citealt{GCK11};
\citealt{Rawle+10}; \citealt{Spolaor+10}; \citealt[herafter
T+10]{Tortora+10CG}, \citealt{Tortora+11CGsim}). Although spectral
features are less ambiguous probes of stellar population
properties than colours are (e.g. \citealt{Spolaor+09,Spolaor+10};
\citealt{Kuntschner+10}; \citealt{Rawle+10}), there are only small
samples available outside the nuclear regions of galaxies.
Fortunately, studies based on colours turn out to provide, on
average, similar age and metallicity estimates to spectral studies
(e.g. T+10), allowing for much larger surveys of radially-extended
stellar populations.

In T+10 we analyzed colour gradients within an effective radius
(\Re) for an extensive sample of low-redshift galaxies based on
SDSS data (\citealt{Blanton05}), including both early-type (ETGs)
and late-type (LTGs) galaxies. We carried out simple stellar
populations analyses, fitting synthetic models to the observed
colours to derive estimated age and metallicity gradients, and
stellar masses (see T+10 and later in the paper for further
details). The main results found can be summarized as follows:
\begin{enumerate}
\item Colour gradients are, on average, negative (i.e. galaxies are redder in the
centre), except for galaxies with lower stellar masses  ($\mst
\lsim 10^{9.3} \, \rm \Msun$) which tend to have null or positive
gradients. At fixed mass, LTGs have steeper gradients than ETGs,
whose colour profiles in general show milder variations .
\item The colour gradients of LTGs decrease systematically with mass while the trend for
ETGs changes near a mass of $\mst \sim 10^{10.3} \, \rm \Msun$:
systematically decreasing and increasing at lower and higher
masses, respectively.
\item The primary driver of the colour gradients, mainly for galaxies with older centers, is metallicity
variations, with age gradients playing a secondary role. For LTGs,
the metallicity gradients tend to be steeper for higher masses,
The same is true for ETGs except that for masses above $\mst \sim
10^{10.3} \, \rm \Msun$, the (negative) gradients are fairly
invariant or slightly increasing with mass.
\item At fixed mass, galaxies with older centres have the shallowest
metallicity and age gradients.
\end{enumerate}

\begin{figure}
\psfig{file=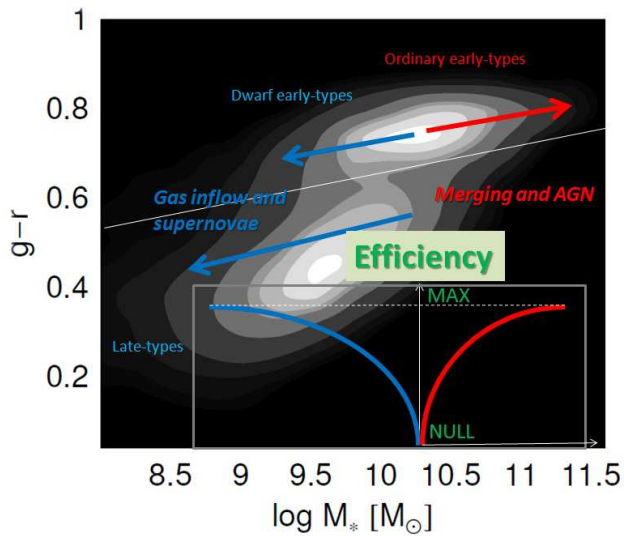, width=0.48\textwidth} \caption{Diagram of
colour versus stellar mass for the local SDSS galaxy sample
analyzed in T+10. The grey-scale contours show the density of data
points, with brighter regions marking higher densities. The thin
white line separates the red sequence and blue cloud. The arrows
and the curves in the inset panel give information about the
efficiency of the phenomena which drive the two fold trend we
discuss in the paper (see T+10 for further details). AGN feedback
and merging are important at high mass with an efficiency that
increases with mass, while SN feedback and gas inflow drive the
galaxy evolution at lower masses, with an efficiency that is
larger at the lowest masses.} \label{fig:fig1}
\end{figure}

We found that these results were generally consistent with others
in the literature (e.g. \citealt{Spolaor+09}; \citealt{Rawle+10})
once different galaxy selection effects were taken into account.

The overall observational picture can be framed within the
predictions of hydrodynamical and chemodynamical simulations of
galaxy formation (\citealt{BS99}; \citealt{Kawata01};
\citealt{KG03}; \citealt{Ko04}; \citealt{Hopkins+09a};
\citealt{Pipino+10}; \citealt{Tortora+11CGsim}) which include
various physical processes that can influence the gradients:
supernovae and AGN feedback, galaxy mergers, interactions with
environment, etc. A fresh and qualitative overview of this
physical scenario is given by Fig. \ref{fig:fig1}, where we show
the isodensity contours for the colour-mass diagram (which we have
already discussed in T+10) and superimpose some lines (in the
inset) and arrows which indicate the efficiency of the highlighted
physical processes. In particular, the observations for high-mass
ETGs seem to generally suggest the influence of mergers and AGN
feedback (\citealt{Ko04}; \citealt{Hopkins+09a};
\citealt{Sijacki+07}), with an efficiency possibly increasing with
mass (see red arrows and lines in the inset panel). On the other
hand, the population gradients of LTGs and less massive ETGs may
be mainly driven by infall and supernovae feedback
(\citealt{Larson74, Larson75}; \citealt{Kawata01}; \citealt{KG03};
\citealt{Ko04}; \citealt{Pipino+10}; \citealt{Tortora+11CGsim}),
which are more efficient at very low masses (see blue lines and
arrows).

In this paper we consider the implications of the observed colour,
metallicity and age gradients from T+10 for variations of stellar
{\it mass-to-light ratio} $M/L$ with radius.  This issue is
relevant for understanding the detailed distribution of mass
within galaxies:  both the stellar mass and the dark matter, whose
distribution is inferred by substracting the stellar mass from the
total dynamical mass. There has been scarcely any literature to
date that focussed on this topic, except for a study in spiral
galaxies by \cite{PS10}.

As a first approximation, one could use single-colour
semi-empirical models such as those of \cite{BdJ01} and \citet[B03
hereafter]{Bell+03} to go directly from colour gradients to \ML\
gradients. However, we wish to make use of multiple colours, to
allow for departures of individual galaxies from mean trends, to
explore the systematics thoroughly, and to trace the dependence on
stellar parameters, like age and metallicity.

This paper is organized as follows. The sample
and the spectral analysis are discussed in \S \ref{sec:sample}, in
where we also show the ability of our fitting procedure
to recover the correct \ML\ gradients, and we examine possible
systematics. The results are shown in \S \ref{sec:results}, while
\S \ref{sec:conclusions} is devoted to a discussion of the
physical implications and conclusions.

\section{Data and spectral analysis}\label{sec:sample}

Our database consists of $50\,000$ low redshift ($0.0033 \leq z
\leq 0.05$) galaxies in the NYU Value-Added Galaxy Catalog
extracted from SDSS DR4 (\citealt[hereafter
B05]{Blanton05})\footnote{The low redshift NYU-VAG catalogue is
available at: {\tt http://sdss.physics.nyu.edu/vagc/lowz.html}.},
recently analyzed in T+10\footnote{See T+10 for further details
about sample selection, incompleteness and biases.}. Following
T+10, we have identified ETGs as those systems with a r-band S$\rm
\acute{e}$rsic index satisfying the condition $2.5 \leq n \leq
5.5$ and with a concentration index $C > 2.6$. The final ETG
sample consists of $10\,508$ galaxies. Of the remaining entries of
our database, $\sim 27\,800$ are LTGs, defined as  objects with $C
\leq 2.6$, $n \leq 2.5$, and $\sigc\leq 150~\rm km/s$.

\begin{table*}
\centering \caption{Recovery of stellar population parameters from
Monte Carlo simulations. A purely random sample of galaxies is
generated, leaving the ages and metallicities free to vary across
the whole range. We list the median error  and the 25-75 percent
scatter in the gradient $\Delta(X)=X_{\rm fit}-X_{\rm in}$, where
$X_{\rm in}$ and $X_{\rm fit}$ are the input and fitted gradients
with $X = \gML$, $\gage$ and $\gZ$. The data are divided in three
subsamples, on the basis of the input \ML s at $R_{2}$. Different
initial perturbations $\delta$ and the results (1) with optical
only ($ugriz$) (2) optical and near-IR ($ugrizJHKs$), and (3)
optical and UV ($ugrizFUVNUV$) are shown.}\label{tab:tab1}
\resizebox{17.5cm}{!}{
\begin{tabular}{lccccccccc}
\hline
 & \multicolumn{3}{c}{$\log \Yst_{in,2} \leq 0$} & \multicolumn{3}{c}{$0< \log \Yst_{in,2} \leq 0.4$} & \multicolumn{3}{c}{$\log \Yst_{in,2} >
 0.4$} \\
 & $ugriz$ & $ugrizJHKs$ & $ugrizFUVNUV$ & $ugriz$ & $ugrizJHKs$ & $ugrizFUVNUV$ & $ugriz$ & $ugrizJHKs$ & $ugrizFUVNUV$ \\
\hline
& \multicolumn{9}{c}{$\gML$} \\
\hline
$\delta = 0.01$  & $0_{-0.02}^{+0.02}$ & $0_{-0.02}^{+0.02}$ & $0_{-0.02}^{+0.02}$ & $0_{-0.03}^{+0.04}$ & $0_{-0.02}^{+0.02}$ & $0_{-0.02}^{+0.02}$ & $0_{-0.04}^{+0.03}$ & $0_{-0.03}^{+0.02}$ & $0_{-0.02}^{+0.02}$\\
$\delta = 0.03$  & $0_{-0.05}^{+0.04}$ & $0_{-0.03}^{+0.03}$ & $0_{-0.02}^{+0.02}$ & $0.01_{-0.06}^{+0.10}$ & $0_{-0.03}^{+0.03}$ & $0_{-0.03}^{+0.03}$ & $-0.01_{-0.07}^{+0.07}$ & $0_{-0.04}^{+0.04}$ & $0_{-0.02}^{+0.02}$\\
$\delta = 0.05$  & $0_{-0.07}^{+0.06}$ & $0_{-0.04}^{+0.04}$ & $0_{-0.03}^{+0.02}$ & $0.02_{-0.10}^{+0.12}$ & $0.01_{-0.05}^{+0.05}$ & $0_{-0.04}^{+0.04}$ & $-0.01_{-0.10}^{+0.09}$ & $0_{-0.06}^{+0.06}$ & $0_{-0.03}^{+0.03}$\\

\hline

& \multicolumn{9}{c}{$\gage$} \\
\hline
$\delta = 0.01$  & $0_{-0.04}^{+0.04}$ & $0_{-0.02}^{+0.03}$ & $0_{-0.02}^{+0.02}$ & $0.01_{-0.05}^{+0.06}$ & $0_{-0.03}^{+0.03}$ & $0_{-0.03}^{+0.03}$ & $0.01_{-0.06}^{+0.06}$ & $0_{-0.04}^{+0.04}$ & $0_{-0.03}^{+0.03}$\\
$\delta = 0.03$  & $0_{-0.08}^{+0.08}$ & $0_{-0.04}^{+0.05}$ & $0_{-0.03}^{+0.03}$ & $0.02_{-0.12}^{+0.16}$ & $0_{-0.05}^{+0.05}$ & $0_{-0.04}^{+0.04}$ & $-0.02_{-0.12}^{+0.11}$ & $0_{-0.06}^{+0.07}$ & $0_{-0.04}^{+0.03}$\\
$\delta = 0.05$  & $-0.01_{-0.12}^{+0.13}$ & $0_{-0.07}^{+0.07}$ & $0_{-0.04}^{+0.04}$ & $0.04_{-0.17}^{+0.21}$ & $0.01_{-0.08}^{+0.08}$ & $0.01_{-0.06}^{+0.06}$ & $-0.02_{-0.16}^{+0.15}$ & $-0.01_{-0.09}^{+0.09}$ & $0_{-0.05}^{+0.05}$\\

\hline

& \multicolumn{9}{c}{$\gZ$} \\
\hline
$\delta = 0.01$  & $0_{-0.05}^{+0.06}$ & $0_{-0.03}^{+0.03}$ & $0_{-0.03}^{+0.03}$ & $-0.01_{-0.07}^{+0.06}$ & $0_{-0.03}^{+0.03}$ & $0_{-0.04}^{+0.03}$ & $0.00_{-0.05}^{+0.05}$ & $0_{-0.03}^{+0.03}$ & $0_{-0.03}^{+0.03}$\\
$\delta = 0.03$  & $0_{-0.12}^{+0.12}$ & $0_{-0.06}^{+0.05}$ & $0_{-0.04}^{+0.04}$ & $-0.02_{-0.18}^{+0.16}$ & $0_{-0.05}^{+0.05}$ & $0_{-0.04}^{+0.04}$ & $0.02_{-0.10}^{+0.12}$ & $0_{-0.05}^{+0.05}$ & $0_{-0.04}^{+0.04}$\\
$\delta = 0.05$  & $0.01_{-0.20}^{+0.19}$ & $0_{-0.08}^{+0.09}$ & $0_{-0.05}^{+0.05}$ & $-0.03_{-0.26}^{+0.22}$ & $0.01_{-0.08}^{+0.09}$ & $-0.01_{-0.06}^{+0.06}$ & $0.02_{-0.15}^{+0.16}$ & $-0.01_{-0.07}^{+0.07}$ & $0_{-0.05}^{+0.05}$\\

\hline

\end{tabular} }
\end{table*}

\begin{table*}
\centering \caption{Median errors as in Table \ref{tab:tab1}, but
for $\delta = 0.05$. Ages and metallicities are generated to have
average \gage\ and \gZ\ around fixed values.}\label{tab:tab1bis}
\resizebox{17.5cm}{!}{
\begin{tabular}{lccccccccc}
\hline
 & \multicolumn{3}{c}{$\log \Yst_{in,2} \leq 0$} & \multicolumn{3}{c}{$0< \log \Yst_{in,2} \leq 0.4$} & \multicolumn{3}{c}{$\log \Yst_{in,2} >
 0.4$} \\
 & $ugriz$ & $ugrizJHKs$ & $ugrizFUVNUV$ & $ugriz$ & $ugrizJHKs$ & $ugrizFUVNUV$ & $ugriz$ & $ugrizJHKs$ & $ugrizFUVNUV$ \\
\hline
& \multicolumn{9}{c}{$\gML$} \\
\hline

$\gage \sim 0.09$, $\gZ \sim -0.32$  & $-0.01_{-0.09}^{+0.10}$ & $0.01_{-0.06}^{+0.07}$ & $0_{-0.03}^{+0.03}$ & $-0.01_{-0.12}^{+0.12}$ & $0_{-0.08}^{+0.07}$ & $0_{-0.05}^{+0.04}$ & $-0.06_{-0.14}^{+0.11}$ & $-0.01_{-0.08}^{+0.06}$ & $-0.01_{-0.04}^{+0.03}$\\
$\gage \sim -0.12$, $\gZ \sim -0.47$  & $-0.02_{-0.10}^{+0.07}$ & $0_{-0.06}^{+0.05}$ & $0_{-0.03}^{+0.03}$ & $-0.03_{-0.12}^{+0.12}$ & $0_{-0.07}^{+0.07}$ & $-0.01_{-0.05}^{+0.04}$ & $-0.07_{-0.14}^{+0.11}$ & $-0.02_{-0.07}^{+0.06}$ & $-0.01_{-0.03}^{+0.03}$\\
$\gage \sim 0.35$, $\gZ \sim -0.08$  & - & - & - & - & - & - & $-0.05_{-0.12}^{+0.09}$ & $-0.01_{-0.07}^{+0.06}$ & $0_{-0.03}^{+0.03}$\\

\hline

& \multicolumn{9}{c}{$\gage$} \\
\hline

$\gage \sim 0.09$, $\gZ \sim -0.32$  & $-0.05_{-0.16}^{+0.20}$ & $0_{-0.10}^{+0.10}$ & $0_{-0.06}^{+0.05}$ & $-0.02_{-0.21}^{+0.21}$ & $0.01_{-0.11}^{+0.11}$ & $-0.01_{-0.08}^{+0.07}$ & $-0.10_{-0.25}^{+0.19}$ & $-0.02_{-0.12}^{+0.10}$ & $-0.01_{-0.06}^{+0.04}$\\
$\gage \sim -0.12$, $\gZ \sim -0.47$  & $-0.03_{-0.16}^{+0.13}$ & $0_{-0.08}^{+0.08}$ & $-0.01_{-0.05}^{+0.05}$ & $-0.05_{-0.20}^{+0.19}$ & $0_{-0.10}^{+0.11}$ & $-0.01_{-0.07}^{+0.06}$ & $-0.13_{-0.24}^{+0.18}$ & $-0.03_{-0.10}^{+0.10}$ & $-0.01_{-0.06}^{+0.05}$\\
$\gage \sim 0.35$, $\gZ \sim -0.08$  & - & - & - & - & - & - & $-0.08_{-0.21}^{+0.16}$ & $-0.02_{-0.11}^{+0.10}$ & $-0.01_{-0.05}^{+0.05}$\\
\hline

& \multicolumn{9}{c}{$\gZ$} \\
\hline

$\gage \sim 0.09$, $\gZ \sim -0.32$  & $0.11_{-0.23}^{+0.17}$ & $0.01_{-0.09}^{+0.08}$ & $0.01_{-0.07}^{+0.08}$ & $0.03_{-0.21}^{+0.24}$ & $0_{-0.09}^{+0.09}$ & $0.01_{-0.06}^{+0.07}$ & $0.09_{-0.18}^{+0.23}$ & $0.02_{-0.09}^{+0.08}$ & $0.02_{-0.06}^{+0.06}$\\
$\gage \sim -0.12$, $\gZ \sim -0.47$  & $0.05_{-0.18}^{+0.19}$ & $0_{-0.10}^{+0.11}$ & $0.01_{-0.05}^{+0.06}$ & $0.05_{-0.18}^{+0.23}$ & $0_{-0.08}^{+0.08}$ & $-0.01_{-0.06}^{+0.07}$ & $0.10_{-0.15}^{+0.22}$ & $0.02_{-0.07}^{+0.08}$ & $0.02_{-0.05}^{+0.06}$\\
$\gage \sim 0.35$, $\gZ \sim -0.08$  & - & - & - & - & - & - & $0.06_{-0.14}^{+0.19}$ & $0.02_{-0.07}^{+0.08}$ & $0_{-0.05}^{+0.05}$\\
\hline

\end{tabular} }
\end{table*}

We have used the multi-band structural parameters given by B05 to
derive the colour profile $(X-Y)(R)$ of each galaxy as the
differences between the (logarithmic) surface brightness
measurements in the two bands, $X$ and $Y$. We have defined the
colour gradient as the angular coefficient of the relation $X-Y$
vs $\log R/\Re$, $\displaystyle \nabla_{X-Y} = \frac{\delta
(X-Y)}{\delta \log (R/R_{\rm eff})}$, measured in mag~dex$^{-1}$
(omitted in the following),  where \Re\ is the r-band effective
radius. By definition, a positive gradient, $\nabla_{X-Y}>0$,
means that a galaxy is redder as $R$ increases, while it is bluer
outward for a negative gradient. The fit of synthetic colours is
performed on the colours at $R_{1} = \Re / 10$ and $R_{2} = \Re$
and on the total integrated colours.

The stellar population properties (age, metallicity and \ML) are
derived by fitting \citet[hereafter BC03]{BC03} ``single burst''
synthetic stellar models to the observed colours. Age and
metallicity are free to vary, and a \cite{Chabrier01} initial mass
function (IMF) is assumed. Stellar parameter gradients are defined
as $\displaystyle \nabla_{W} = \frac{\delta (W)}{\delta \log
(R/R_{\rm eff})}$, where $W = (age, Z, \Yst)$ are the estimated
age, metallicity and \ML. Because of the definitions adopted for
$R_{1}$ and $R_{2}$, an easier and equivalent definition can be
used, in fact, we define $\gage =\log [\rm age_{2}/\rm age_{1}]$,
$\gZ=\log [Z_{2}/ Z_{1}]$ and $\gML=\log [\rm \Yst_{2}/ \rm
\Yst_{1}]$ where (age$_{i}$, $Z_{i}$, $\Yst_{i}$), with $i=1,2$,
are the estimated parameters at $R_{1}$ and $R_{2}$,
respectively\footnote{We use $B$-band \ML, but for simplicity we
omit the subscript $B$.}. In the present paper we will concentrate
on the analysis of the \ML\ gradients, \gML, in terms of stellar
mass and velocity dispersion. The dataset, including main stellar
parameters and gradients, is available at: {\tt
http://www.itp.uzh.ch/$\rm\sim$ctortora/gradient\_data.html}.

Before proceeding, we check for systematic effects on our \gML\
estimates from the stellar population fits. Because of the well
known age-metallicity degeneracy (\citealt{Worthey94},
\citealt{BC03}, \citealt{Gallazzi05}), the stellar population
parameters and the gradients might be biased when using the
optical colours only, as we do now. Widening the wavelength
baseline to include near-infrared (NIR) colours should ameliorate
the degeneracy.  In T+10 we checked age and metallicity inferences
using optical versus optical+NIR constraints, and found little
systematic difference. Here we will carry out a similar analysis
of the \ML\ gradients, using a suite of Monte Carlo simulations.

We  extract some sets of simulated galaxy spectra from our BC03
spectral energy distribution libraries with random, uncorrelated
stellar parameters and gradients. We start from generating data
with fully random ages and metallicities, and then we also analyze
the cases when the stellar parameters are constrained within fixed
intervals, generating not-zero sample average gradients. We
extract the input galaxy colours and add simulated measurement
errors, as randomly extracted steps from the interval ($-
\delta$,$+ \delta$), with $\delta = 0.01, 0.03, 0.05$. We apply
our fitting procedure, searching for the best model in our
synthetic library, which reproduces the colours of each of the
simulated galaxies and then compare the output parameter estimates
to the input model values. We perform the fit (1) using only the
optical SDSS bands $ugriz$ and (2) adding NIR photometry ($J$, $H$
and $Ks$; \citealt{Jarrett+03}) and (3) ultraviolet photometry
(NUV and FUV; \citealt{Martin+05}). We define the median error in
the gradients as $\Delta(X)=X_{\rm fit}-X_{\rm in}$, where $X_{\rm
in}$ and $X_{\rm fit}$ are the input and fitted gradients with $X
= \gML$, $\gage$ and $\gZ$, which we show in Tables \ref{tab:tab1}
and \ref{tab:tab1bis}.

Starting from the purely random case shown in Table
\ref{tab:tab1}, we have found that, adding the NIR or UV data,
both the \ML\ values and their gradients are perfectly recovered
with a very little scatter. With optical data only, the
uncertainties are doubled (around 0.1 in $\Delta(\gML)$) and there
is a small systematic shift in the gradients (of $\log \gML \sim
0.01-0.02$ for the worst case analyzed), but the results are
fairly well recovered with no spurious trends expected. We notice
that the median results are quite independent of the range adopted
for the input $\Yst_{2}$, and only slight differences in the
scatter are found. Although we will be mainly interested in the
trends of \gML\ with mass and velocity dispersion, we have also
checked the systematics in age and metallicity gradients. We find
similar results to the case of \gML\ discussed above, but slightly
larger shifts in the median and larger scatters ($\sim 0.2$ in the
worst cases) when only the optical is used. In Table
\ref{tab:tab1bis}, calculating the sample averages, we start with
the input values $\gage \neq 0$ and $\gZ \neq 0$. Here the median
errors and the scatters increase, but the conclusions discussed
above are qualitatively confirmed. When optical data are used, for
high $\Yst_{2}$ we have the worst discrepancies in the estimated
\ML\ gradients amounting to $\sim -0.05$, $-0.07$. For age and
metallicity gradients, we could underestimate (overestimate) the
age (metallicity) gradients of $\sim 0.1$, which is reduced to
$\sim 0.02$ when near-IR or UV are included.

Thus, in many cases we have found that the median errors are
acceptable and within the typical sample scatter and the
uncertainties due to the adopted stellar population prescription.
We will consider these results as qualitative indications of the
effects on estimated gradients caused by systematics.

\begin{figure*}
\hspace{-1.2cm} \psfig{file=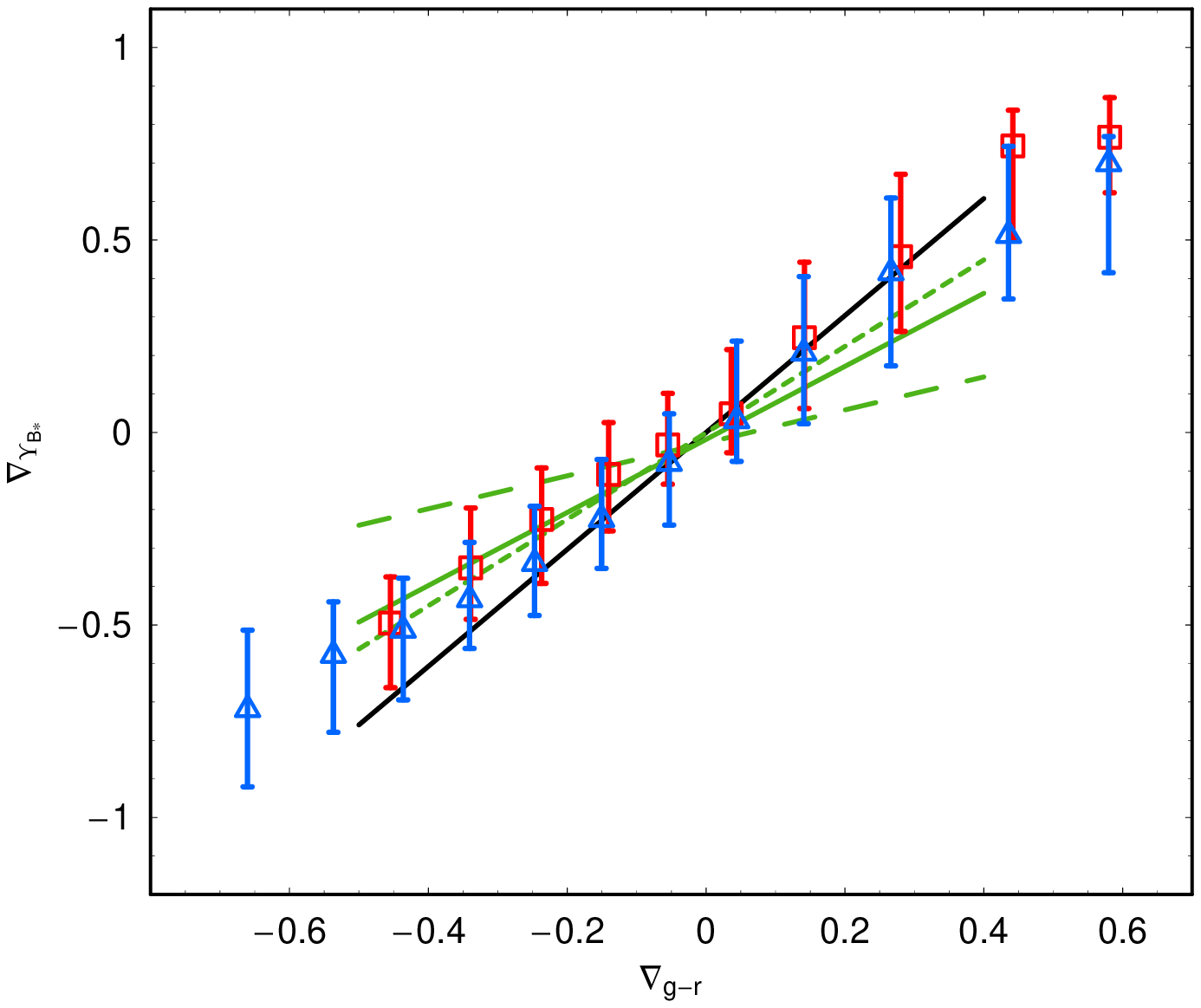, width=0.55\textwidth}
\psfig{file=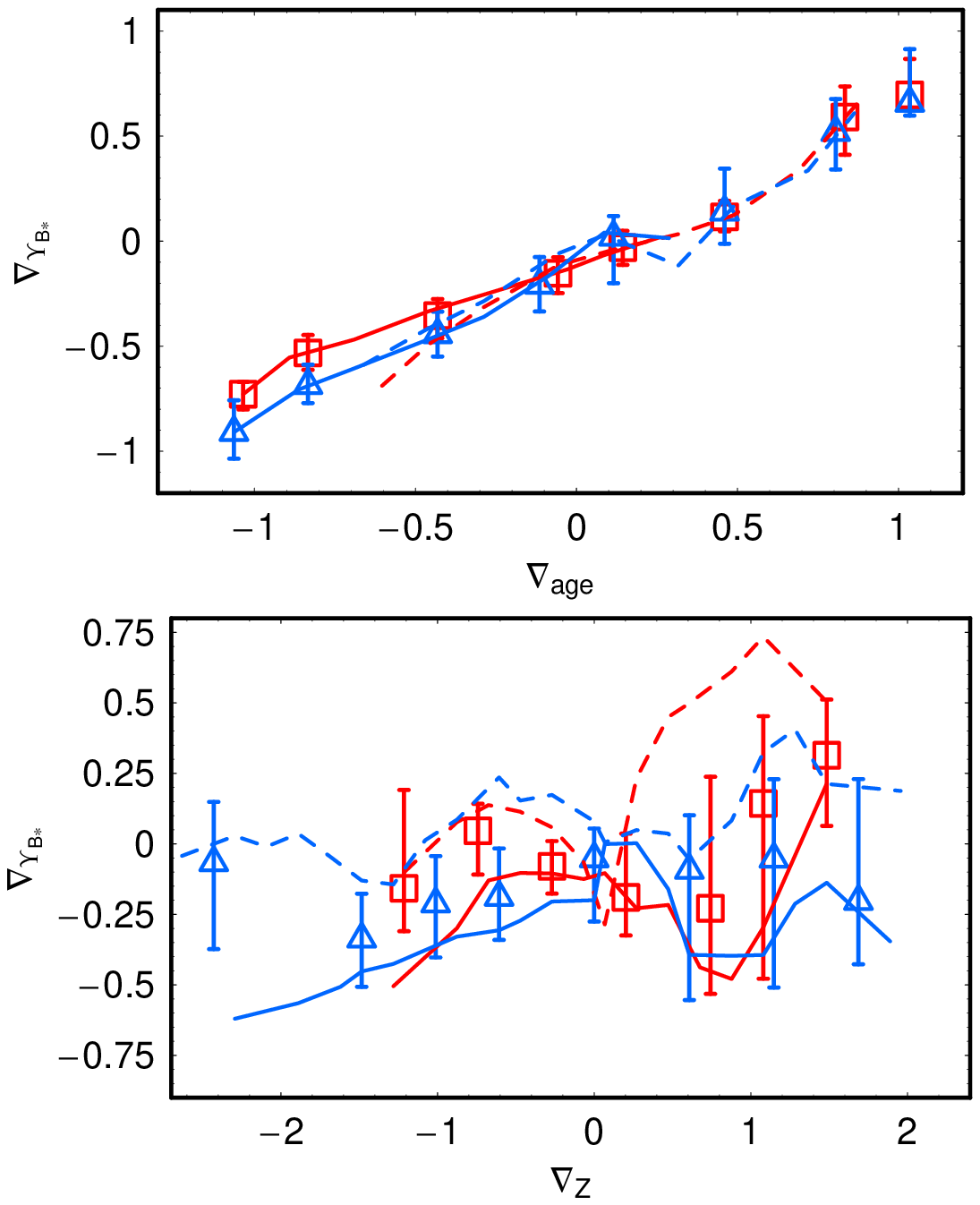, width=0.39\textwidth} \caption{{\it Left}.
\ML\ gradient as a function of $g-r$ gradient for ETGs (red
squares) and LTGs (blue triangles). Medians of the gradients (in
bins of colour gradient) are plotted together with the 25-75 per
cent quantiles, shown as error bars. The lines are for the
gradients obtained using the best fitted \ML-colour relations in
B03. The relationships between the $g$-band \ML\ ($g$-band is
approximatively the same as the $B$-band) and $g-r$ (solid black
line), $u-g$ (long dashed green line), $g-i$ (short dashed green
line) and $g-z$ (green line) are adopted and the recovered \gML\
are plotted as a function of \ggr. {\it Right.} Age (top) and
metallicity (bottom) gradients as a function of $g-r$ colour
gradient. Solid and dashed lines show systems whose centres are
older and younger than 6~Gyr, respectively. \ML\ and colour
gradients are tightly correlated for both ETGs and LTGs, and,
while \gML\ correlates with \gage, no clear trend is observed in
terms of \gZ.} \label{fig:fig2}
\end{figure*}

\section{Results}\label{sec:results}

In this section we discuss first the \ML\ gradients in terms
of colour, age and metallicity gradients and then as a function of
stellar mass and velocity dispersion, paying attention to the role
of central age in the observed correlations.

\subsection{Dependence on colour, age and metallicity gradients}

We start by plotting in Fig. \ref{fig:fig2} the \ML\ gradients as
a function of $g-r$ colour gradient ($\ggr$) for ETGs and LTGs.
The two quantities are strongly and positively correlated and no
statistically significant difference between the morphological
types is found. While such a correlation is not surprising owing
to the well known-correlations between colours and \ML\
(\citealt{BdJ01}; B03), which should apply to all radii thus
producing the relationship between \gML\ and \ggr, there are
important details to consider. \cite{BdJ01} have adopted different
evolution models for LTGs to establish some linear correlations
between integrated total colours and the corresponding \ML\
determining some best fit relations between these two quantities,
which have been proven to work for ETGs as well
(\citealt{Bell+03}). Adopting the solar metallicity models in
\cite{BdJ01} we have found a good agreement with \cite{BC03} or
\cite{Maraston05} synthetic prescriptions in ETGs with high \ML,
while at low \ML\ \cite{BdJ01} overestimates the \ML\ (see
\cite{Tortora2009} for further details). In the left panel of Fig.
\ref{fig:fig2} we compare our \gML-\ggr\ trends with the ones
(solid and dashed lines) derived from the B03 best fitted
relations, finding a good agreement. The black and green lines
plotted in Fig. \ref{fig:fig2} are derived by the relations
between the g-band \ML\ and the colours ($u-g$, $g-r$, $g-i$ and
$g-z$) from B03. We insert in these relations the measured colour
for each galaxy in our sample, deriving the corresponding \ML.
This operation is made for both the radii $R_{1}$ and $R_{2}$,
thus we can calculate the \ML\ gradients. Finally, the plotted
lines are obtained, first calculating medians of \gML\ in bins of
\ggr\ and then fitting a straight line. We see that the best
fitted relations depend on the colours adopted in the colour--\ML\
relations from B+03 to derive the \ML\ gradients. For instance, we
observe a flattening of the \gML --\ggr\ relation when using
colours probing redder wavelengths, i.e. $g-i$ and $g-z$, but a
shallower trend is obtained when $u-g$ is adopted. The agreement
is quite good for moderate colour gradients, while for very steep
positive or negative gradients some departures are observed.

Despite the good agreement, the very simplified approach in
\cite{BdJ01} and B03 has some strong limitations. In fact, a) the
B03 relations hold for integrated quantities, thus the radial
variation of the $M/L$ can be derived if one assumes that the same
relations apply at different apertures and in general, the
zero-point and the slope of these fitted linear relations would
depend on the radius adopted and b) these relations do not take
into account the stellar parameters (e.g. age and metallicity at
each radius) thus they do not allow to catch the full physics
behind these correlations. These shortcomings are overcome by our
multi-colour direct stellar population analysis which connects the
\ML\ gradients self-consistently to the underlying stellar
population properties (such as age and metallicity). This also
allows us to track the physical reasons for the \ML\ variations,
which in turn should be reproduced by galaxy formation models.

Differently from what happens for colour gradients which are
mostly driven from metallicity gradients, from the right column of
Fig. \ref{fig:fig2} it is evident that the \gML--\ggr\ relation is
driven mainly by {\it age}, in both LTGs and ETGs.

\begin{figure*}
\psfig{file=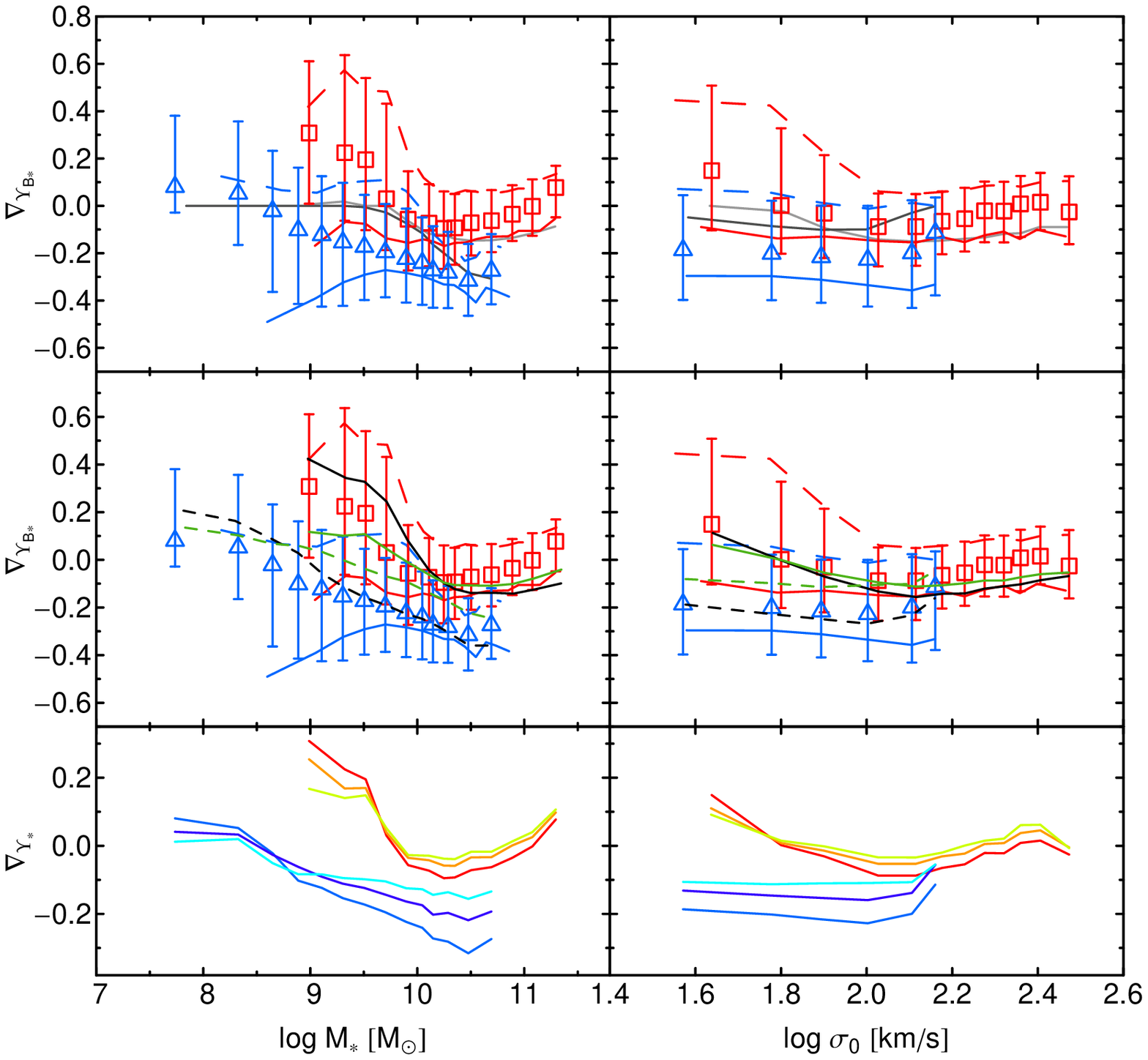, width=0.9\textwidth} \caption{ \ML\
gradients as a function of stellar mass (left) and velocity
dispersion (right). For ETGs a two fold trend in terms of both
\mst\ and \sigc\ is found. LTGs present a steepening of the
gradients with \mst, while they are independent of \sigc. {\it
Top}: Default $B$-band gradients. The symbols are as in Fig.
\ref{fig:fig2}, with solid and dashed lines indicating galaxies
with central $\rm age$ older and younger than $6 \, \rm Gyr$,
respectively. The light and dark grey lines are the results for
ETGs and LTGs when no age gradients are assumed (adopting $\rm age
= 10 \, \rm Gyr$). {\it Middle}: Default $B$-band gradients,
symbols are as panels above. Black and green lines are the medians
obtained using the $\ML-(g-r)$ and $\ML-(u-g)$ relations from B03,
continue and dashed are for ETGs and LTGs. {\it Bottom}: Average
gradients using different bandpasses: $B$ (red and blue lines for
ETGs and LTGs, respectively), $V$ (orange and violet) and $i$
(yellow and cyan).} \label{fig:fig3}
\end{figure*}

\subsection{Trends with stellar mass and velocity dispersion}

The trends of \gML\ with stellar mass and velocity dispersion are
shown in the top panels of Fig. \ref{fig:fig3}. As with the trend
of colour and metallicity gradients with mass (see comments in \S
\ref{sec:intro} also T+10), we find that LTGs have a decreasing
trend with mass and show \gML$>0$ for $\log \mst/\Msun \lsim 9$
and \gML$<0$ at larger masses. In contrast, ETGs have positive
$\gML \sim 0.25$ at low masses, which decreases down to a minimum
of $\gML \sim -0.1$ at $\log \mst/\Msun \sim 10.3$ and then invert
the trend, with the most massive ETGs having gradients around zero
or slight positive. When plotted as a function of velocity
dispersion, the \gML\ run with $\sigc$ is generally flatter, with
LTGs having no trend and ETGs showing a two-fold trend with a
minimum around $\log \sigc=2.1$ km/s and a mild increase toward
the low and high $\sigc$ sides, always staying consistent with
\gML$\lsim 0$.

Since galaxies are morphologically selected by using criteria on
concentration and S$\rm \acute{e}$rsic index, the steeper
gradients found for LTGs when compared with ETGs indicate a
sequence in terms of such structural parameters. In fact, at fixed
mass, galaxies with lower $C$ or $n$ have steeper colour and \ML\
gradients.

As observed for the stellar population gradients (see T+10), the
central age is responsible for much of the scatter of the trends
versus mass and velocity dispersion, which is illustrated by the
median distribution for centrally old ($> 6 \, \rm Gyr$) and young
($< 6 \, \rm Gyr$) galaxies in Fig. \ref{fig:fig3}. In particular,
young LTGs have almost null gradients, while the oldest ones have
very steep values of $\gML \sim -0.3$ to $-0.4$. ETGs show a
similar but offset effect.  Older ETGs have slightly negative
values of $\sim -0.1$, while the younger ones have steep positive
$\gML \sim 0.5$ at very low masses, decreasing to $\sim 0.1$ at
larger masses.

As a comparison, we also show in Fig.~\ref{fig:fig3} (top panels)
the results when galaxy colours are fitted to a synthetic spectral
model which has no age gradient (fixed $\rm age = 10 \, \rm Gyr$).
In this case, LTGs have \gML\ which are null up to $\log
\mst/\Msun \sim 9.5$, and decrease at larger masses, while
gradients in ETGs show a milder trend when compared with the
reference fit. When plotted versus \sigc\, both LTGs and ETGs have
similar \gML{}~$\sim -0.1$ for all velocity dispersions. Note that
these results are consistent with what was found for the older
galaxies in the reference fit. We also compare our results with
g-band \ML s derived from B03. As an example we plot the median
lines derived using the $\ML-(g-r)$ and $\ML-(u-g)$ relations for
both ETGs and LTGs. The trends with \mst\ and velocity dispersion
are qualitatively unchanged, but some differences emerge since the
\ML s from each B03 relation are derived using different colours,
which probe different wavelength spectral regions. The \ML s
obtained using the $g-r$ colours seems to resemble our trends for
LTGs, while the agreement is poorer when $u-g$ is used. For ETGs
our trend is between the two B03 results at low mass, while the
\ML\ from $u-g$ is a little bit better at large mass.

We have also checked the effect of the waveband adopted for the
\ML\ ratios in the bottom panels of Fig. \ref{fig:fig3}. Here, the
$B$-band \ML\ gradients are compared to the ones in $V$- and
$i$-band. Except for the most massive ETGs, in general the \ML\
gradients tend to become shallower (closer to zero) when using
redder bands. Also, the trends with mass and velocity dispersion
become weaker. Thus, redder bands, which are less affected by dust
extinction and trace the older and metal-richer stellar
populations in galaxies, tend to be less dependent on the
existence of gradients in colour and stellar population
parameters. Extending such analyzing into the NIR bands should
find \ML\ gradients increasingly approaching zero\footnote{ We
prefer to limit the analysis to the wavelength coverage available
for our dataset, avoiding extrapolations to near-IR wavebands.}.

From the preceding results, we can see that although the age
gradient is generally the driving parameter behind the \ML\
gradient, this conclusion can vary depending on galaxy type, mass
and central age. We summarize the gradients for different galaxy
subsamples in Table \ref{tab:tab2}, noting also which of the
stellar parameters drives the \ML\ gradient.

\begin{table*}
\centering \caption{\ML, age and metallicity gradients for
subpopulations of galaxy type, stellar mass and central age. In
(parentheses) we note whether the age or metallicity gradient
drives the \ML\ gradient.}\label{tab:tab2}
\begin{tabular}{lll}
\hline \hline
LTGs & $\log\mst/\Msun\lsim9$  & $\log\mst/\Msun\gsim9$  \\
 \hline
Young & $\gML > 0$; $\gage > 0$; $\gZ \lsim 0$& $\gML < 0$; $\gage > 0$; $\gZ < 0$\\
 & (age) & (metallicity) \\
 \hline
Old & $\gML < 0$; $\gage < 0$; $\gZ \lsim 0$& $\gML < 0$; $\gage \lsim 0$; $\gZ < 0$ \\

 & (age) & (metallicity)\\
\hline \hline
ETGs & $\log\mst/\Msun\lsim10$  & $\log\mst/\Msun\gsim10$  \\
\hline
Young & $\gML > 0$; $\gage > 0$; $\gZ \gsim 0$& $\gML > 0$; $\gage > 0$; $\gZ < 0$\\
 & (age) & (age)\\
 \hline
Old & $\gML < 0$; $\gage \lsim 0$; $\gZ \gsim 0$& $\gML < 0$; $\gage \sim 0$; $\gZ < 0$ \\
& (age) & (metallicity)\\
\hline \hline
\end{tabular}
\end{table*}

\section{Discussions and conclusions}\label{sec:conclusions}

In this paper we have investigated the correlation between \ML\
gradients and stellar mass or velocity dispersion for a large
sample of local galaxies from SDSS. The gradients \gML\ have been
derived by the fitting of synthetic spectral models to observed
optical colours. We have used Monte Carlo simulations to check
that the optical-only wavelength baseline should not dramatically
affect the results compared to including near-infrared and
ultraviolet data. This conclusion is encouraging in view of future
studies of higher redshift galaxies which are generally observed
in rest-frame optical bands. Anyway, future analysis in local
samples, relying on near-IR and/or UV data would be able to give
further details about the degeneracies and systematics affecting
stellar fitting.

We have found that there exists a tight positive correlation
between \gML\ and colour gradients which is not surprising because
of the well known \ML\--colour correlations (e.g.,
\citealt{BdJ01}; B03). We have found that \ML\ is a decreasing
function of stellar mass in LTGs, while a two-fold trend is found
for ETGs (for both stellar mass and velocity dispersion), in
agreement with the trends of colour gradients discussed in T+10.
Similar results are obtained using the \ML\--colour relations from
B03, which, although easy to use, do not give information about
stellar population parameters. Thus, the novelty of our approach
resides in our ability to check how \gML\ gradients are driven by
variations in metallicity or age. In general, we have found that
age is most important, but that metallicity becomes more important
in many high-mass galaxies.

This picture does not change dramatically if \ML\ gradients are
calculated in redder bands ($V$- and $i$-bands rather than $B$).
In these cases \gML\ are shallower and the trends with stellar
mass and velocity dispersion become milder (Fig. \ref{fig:fig3}).
These results suggest that \ML\ estimates from red or infrared
bands would not depend on galactocentric radius, since the
radiation emitted by stellar populations in these spectral regions
is more homogenous.

This is the first work to derive \ML\ gradients 1) from
observations of a large galaxy sample and 2) using a direct
stellar population analysis approach. The only point of comparison
is the work by \cite{PS10} in a sample of spiral galaxies, where
negative \ML\ gradients were found to be important in modelling
rotation curves.

We may explain some of our observed trends through physical
mechanisms as follow. Starting from low-mass systems,
centrally-young LTGs have age increasing with radius while
metallicity decreases, which can be due to the presence of a gas
inflow which allows the formation of younger stars in the centre
with a smaller \ML. Centrally-old LTGs have age decreasing with
radius which implies recent outer star formation with a similar
metallicity to the centre. Here the central star formation might
have been quenched by SNe which have expelled metals to larger
radii (where the younger stars drive a smaller \ML).
Centrally-young ETGs have positive age, metallicity and \ML\
gradients, and may hosting expanding shells (\citealt{Mori+97}).
Centrally-old ETGs show mild negative age gradients and very
shallow positive metallicity gradients, similarly to the old LTGs
thus again suggesting SNe having played a role.

The more massive centrally-young LTGs have decreasing \ML\ with
radius, but also strong negative \gZ, thus the new stars formed
inside the core (\gage$>0$) have higher metallicity and \ML. This
might be due to recycled gas from SNe or AGN which is falls back
to the centre due to the deep potential wells and produces a
rejuvenated stellar population. In centrally-young ETGs, on the
other hand, the younger stars formed later in the centres (i.e.
with positive age gradients) and with higher metallicities,
possibly because of a central burst of star formation in ``wet''
mergers (e.g. \citealt{RJ04}; \citealt{Hopkins+09a};
\citealt{Hopkins+10}; T+10). Centrally-old LTGs and ETGs show very
similar behaviors, which seem compatible with merging events that
largely wash out any age gradients (and the \gZ\ in ETGs), and may
produce coeval metallicities in the outer regions of the LTGs. In
both cases the net effect is that metallicity is the main driver
of the strong negative \gML\ in LTGs and of the weak negative
\gML\ of the massive ETGs (in any). Thus, the physical mechanisms
that can account for the observations outlined in Tab.
\ref{tab:tab2} seem consistent with the evolutionary scenario
discussed in \S\ref{sec:intro}.

In particular, the two-fold trend of ETGs in Fig. \ref{fig:fig3},
with a minimum around $\log\mst/\Msun=10.3$ is mirrored by the
colour and stellar population gradients as in Fig. \ref{fig:fig1}
(and T+10) and found also in other works (\citealt{Rawle+10};
\citealt{Spolaor+09, Spolaor+10}). The characteristic mass agrees
well with the typical mass scale break for star forming and
passive systems (\citealt{Kauffmann2003}), for ``bright'' and
``ordinary'' galaxies (\citealt{Capaccioli92a}; \citealt{GG03};
\citealt{Graham+03}; \citealt{Trujillo+04};
\citealt{Kormendy+09}), or for cold and hot flow (shock heating;
\citealt{dek_birn06}).

Following the main conclusions in T+10, quasi-monolithic collapse
and supernovae feedback could be the main drivers of the
decreasing \gML-\mst\ trends in LTGs and low mass ETGs
(\citealt{Larson74}; \citealt{Kawata01}, \citealt{KG03}), while
the increasing trends in massive ETGs could be explained by the
heightened importance of mergers and AGN feedback (\citealt{Ko04};
\citealt{Sijacki+07}; \citealt{Hopkins+09a}).

\ML\ gradients are expected from simulations of galaxy formation
and evolution. E.g. \cite{BP99} found in chemo-dynamical
simulations of spiral galaxies that the central regions ($< 12 \,
\rm kpc$) have steep negative \ML\ gradients, becoming flatter at
larger radii. High-redshift ($z\sim 2$) merger remnants simulated
by \cite{Wuyts+10} showed time-dependent negative \ML\ as a
consequence of redder cores in redder galaxies (similarly to
observations of local galaxies in T+10). \ML\ gradients were also
predicted for merger remnants by \cite{Hopkins+10}, who stressed
that the effective radius derived from the $B$-band light would
increase with time (by $\sim$~10\%) simply because of stellar
population gradients. This effect is important for conclusions
about stellar densities and related dynamical implications (e.g.
for the fundamental plane). Similarly \citet{van Dokkum08} argued
that the existence of such gradients can imply that the observed
quiescent galaxies at large redshift may be of an order or
magnitude more dense than previously found from observations.

The \ML\ gradients present a general complication for estimates of
the dark matter distribution in galaxies. \citet{PS10} found that
for spiral galaxies, including \ML\ gradients would change the
fits to circular velocity curves but not produce strong effects on
the inferred dark matter density parameters. To our knowledge,
this issue has not been addressed in the literature for early-type
galaxies, and it is beyond the scope of the present paper to do so
in any detail. In particular, we have studied the stellar
population gradients only inside 1~\Re, while the gradients out to
$\sim$~5~\Re\ could be important for dark matter analyses (e.g.
\citealt{2005MNRAS.357..691N, Napolitano+09, Napolitano+11}).
Assuming that the gradients do not change radically outside
$\sim$~1~\Re, for the massive ETGs generally studied with lensing
or dynamics ($\sig_0 \gsim$~150~km~s$^{-1}$), we would expect the
stellar \ML\ gradients to be mild (see Fig.~\ref{fig:fig3}), and
to not strongly affect the dark matter estimates.

A final issue is that until now, we have assumed that the IMF is
invariant. If there are galaxy-to-galaxy variations in the IMF
(see e.g.
\citealt{2010Natur.468..940V,vDC11,2011arXiv1104.2379G}), then the
stellar \ML\ values are changed only by a constant multiplicative
value, and the implied gradients remain unchanged. However, IMF
variations if real would then likely exist {\it within} galaxies
owing to variations in stellar populations. Indeed, there have
been many suggestions that stars forming at early times had a
non-standard IMF (e.g.
\citealt{2005MNRAS.359..211L,2007MNRAS.374L..29K,van
Dokkum08,2008MNRAS.385..147D,Holden10}).

One study of nearby ETGs suggested that systems with older stars
within 1~\Re\ have ``lighter'' (closer to Chabrier than Salpeter)
IMFs \citep{NRT10}. This would imply that positive and negative
age gradients would decrease and increase the stellar \ML\
gradients, respectively. Revisiting Fig. \ref{fig:fig3}, the
gradients for the centrally-young objects would decrease and the
centrally old objects would increase, possibly reducing and even
eliminating the age-related scatter in the gradient trends.

Another scenario would be if the central regions of massive ETGs
have ``higher-mass'' IMFs than standard (\citealt{vDC11}), while
the outer regions have accreted galaxies with more ``normal'' IMFs
(e.g. \citealt{Kroupa01} or \citealt{Chabrier03}). This would
decrease these galaxies' stellar \ML\ gradients, although it is
not clear how this effect would relate to age and metallicity
trends.

In future analyses, we plan to analyze stellar \ML\ profiles out
to a few \Re, and analyze the impact of \ML\ gradients on dark
matter profile inferences, which may be compared to $\Lambda$CDM
(e.g. \citealt{NFW96}). We will also consider the (probably small)
adjustments implied for central dark matter inferences (e.g.
\citealt{Tortora10lensing}). Dust effects will further be
examined: dust is not important for metallicity gradients (see
T+10) but could affect age and thereby \ML\ gradients,
particularly for intermediate-mass ETGs.

\section*{Acknowledgments}

We thank the referee for his suggestions which helped to improve
the paper. CT was supported by the Swiss National Science
Foundation. AJR was supported by National Science Foundation
Grants AST-0808099 and AST-0909237.

\end{document}